&latex209
\documentstyle[psbox, PASJadd]{PASJ95}
\newcommand{\vol}{}
\newcommand{\no}{}
\newcommand{\lett}{}
\newcommand{\spage}{}
\newcommand{\rdate}{2000 June 26}
\newcommand{\adate}{2000 October 24}

\begin{document}

\title{The X-Ray Structure of the Supernova Remnant 3C 400.2}
\author{Kumi {\sc Yoshita}, Hiroshi {\sc Tsunemi}, Emi {\sc Miyata},
and Koji {\sc Mori}\\
{\it Department of Earth and Space Science, Graduate School of
Science, Osaka University} \\
{\it 1-1 Machikaneyama, Toyonaka, Osaka 560-0043}\\
{\it E-mail(KY): kyoshita@ess.sci.osaka-u.ac.jp}}

\abst{
We present here the results of an X-ray study of the supernova remnant
3C 400.2 (G53.6$-$2.2) using the ASCA data.
3C 400.2 has an unusual morphology at radio wavelengths, suggesting two
SNRs superposed along the same line of sight,
whereas its X-ray emission 
is known to be centrally peaked.
We investigated the X-ray spectral variation across the remnant
using the ASCA GIS and the ROSAT PSPC data.
The X-ray spectra can be well fitted by thin thermal plasma models.
However, there is no significant variation in
the temperature and the ionization parameter across the remnant.
We conclude that it is a single SNR rather than
two overlapping SNRs.
The centrally peaked X-ray morphology and
the thin thermal emission with nearly cosmic abundances
indicate that 3C 400.2 belongs to a class of ``mixed-morphology SNRs''.
We found that the physical parameters of 3C 400.2 are similar to those of
other mixed-morphology SNRs. 
The morphology of 3C 400.2 can be explained 
by a supernova explosion occurring near to the edge of an interstellar cloud.
}

\kword{ISM: individual (3C 400.2) --- supernova remnants ---
X-rays: ISM}

\maketitle
\thispagestyle{headings}

\section{Introduction} 

The supernova remnant (SNR) 3C 400.2 is known to have 
an unusual morphology in the radio band.
With the VLA observations at 327.5 and 1465 MHz, Dubner (1994) reported
two circular shells that overlap each other, from which
one might imagine two interacting SNRs.
The larger shell has a diameter of 22$^{\prime}$ centered near
($\alpha$, $\delta$)(J2000) = 
($19^{\rm h}38^{\rm m}53^{\rm s}$,
$17^\circ13^{\prime}$).
The other shell is 14$^{\prime}$ in diameter, whose center
is almost at the northwestern edge of the larger shell,
($\alpha$, $\delta$)(J2000) = 
($19^{\rm h}38^{\rm m}09^{\rm s}$,
$17^\circ18^{\prime}$).
Goss et al.\ (1975) found that the spectral index in radio flux, $S$,  
has no difference between two shells with $\alpha = -0.62 \pm 0.04$,
where $S = \nu^{\alpha}$.

The X-ray emission from 3C 400.2 was first detected with HEAO 1
(Agrawal et al.\ 1983). The Einstein IPC
observation showed that the X-ray emission fills the interior of
the radio shell (Long et al.\ 1991).
This leads to the fact that 3C 400.2 belongs to a group of SNRs which have
a limb-brightened radio and centrally peaked X-ray morphology.
The X-ray peak is located in the region where the radio shells overlap
each other.
The ROSAT PSPC spectrum could be fitted by a thin thermal plasma model
(Saken et al.\ 1995).
They found a small increase of the hardness ratio in
the X-ray bright region,
although the values of the column density and the electron temperature
were highly correlated.

Optically, 3C 400.2 can be described as an incomplete shell
structure of
diffuse filaments with a diameter of 16$^{\prime}$ (Winkler et al.\ 1993).
It is smaller than the radio shell.
Although the optical shell is located at a region with high 
surface brightness on the western side of the radio shells, 
there is little correlation among the three images at
the optical, radio, and X-ray wavelengths.

The distance to 3C 400.2 remains uncertain.
Case and Bhattacharya (1998) derived a distance of 5.0~kpc 
from the systemic velocity of the optical filaments
obtained in Rosado et al.\ (1983), using a more modern rotation curve.
On the other hand,
Giacani et al.\ (1998) estimated the distance to be 2.3~kpc 
based on an observation of the H\,{\small I} cloud toward 3C 400.2.
Other distance estimates are based on the
highly uncertain method of 
the $\Sigma$--{\it D} relation and vary from 3.8 to 6.9~kpc
(Clark, Caswell 1976; Caswell, Lerche 1979; Milne 1979;
Allakhverdiev 1983; Dubner et al.\ 1994).

\begin{table*}[t]
\begin{center}
Table~1.\hspace{4pt}Summary of ASCA and ROSAT observations.\\
\end{center}
 \begin{tabular*}{\textwidth}{@{\hspace{\tabcolsep}
\extracolsep{\fill}}lllccc}
  \hline\hline \\ [-6pt]
   Mission &Sequence & Date of &\multicolumn{2}{c}{Field center} &
  Effective exposure \\
   & number & observation & $\alpha$(J2000) & $\delta$(J2000) &
  (ks) \\ [4pt]\hline\\[-6pt]
  ASCA & 54024000 & 1996 Apr 13-14&
  $19^{\rm h}38^{\rm m}34^{\rm s}$ &
  $17^\circ23^{\prime}42^{\prime\prime}$ & 
  19 (GIS) \\
  ROSAT & RP500190N00 & 1992 Oct 4-10 &
  $19^{\rm h}38^{\rm m}24^{\rm s}$ &
  $17^\circ19^{\prime}48^{\prime\prime}$ & 
  3.6 \\ [4pt]
  \hline
 \end{tabular*}
\end{table*}

Only a few examples of interacting or overlapping SNRs
have been reported.
Among these, two cases have been found 
in the Large Magellanic Cloud.
One is DEM L316, which shows two overlapping shells
at radio and optical wavelengths (Williams et al.\ 1997).
The overlapping area is smaller than that of 3C 400.2,
and two shells are clearly separated.
Unlike 3C 400.2, there are two circular regions
of the X-ray emission, corresponding to the two shells.
The recent ASCA observation of DEM L316
found that the X-ray properties in two shells
have different electron temperatures,
ionization parameters, and elemental abundances
(Nishiuchi et al.\ 2001).
They concluded that DEM L316 is surely two SNRs.
The other is the SNR N186D, which is believed to be interacting with
the N186E region (Rosado et al.\ 1990).
They suggest that N186E is an old SNR which is ionized by internal
stars.
No X-ray study of N186D has been reported (Williams et al.\ 1999).

In our galaxy, Yoshita et al.\ (2000) found that
the X-ray emission extends over
the radio shell of the known radio SNR G69.7+1.0.
They suggest that the X-ray emission comes from
a different SNR, G69.4+1.2, which has
a shell of 1$^{\circ}$ diameter at radio and optical wavelengths.
In this case,
the large shell of G69.4+1.2 surrounds
a radio shell of G69.7+1.0 with 16$^{\prime}$ diameter,
and are overlapping.
A similar situation has been found
in the region of CTB 1 (G116.9+0.2) by Craig et al.\ (1997).
They suggest that the limb-like X-ray emission surrounding 
CTB 1 may be a previously unknown SNR, G117.7+0.6.

In this paper, we report on an X-ray study of 3C 400.2
using the ASCA data.
With only the ROSAT PSPC data,
the obtained parameters were the column density and 
the electron temperature, which were strongly coupled.
The high-energy resolution of ASCA allowed us
to determine the elemental abundances
as well as the ionization parameter of the plasma.
Although the spatial resolution of ASCA is about 3$^{\prime}$,
it is possible to search for the spectral variation across 3C 400.2,
owing to its angular size of 33$^{\prime}$ $\times$ 28$^{\prime}$.
We also used the ROSAT PSPC data to cover the low-energy band. 

\section{Observations of 3C 400.2}
 \begin{figure*}[t]
   \centerline{\psbox[xsize=1#1]{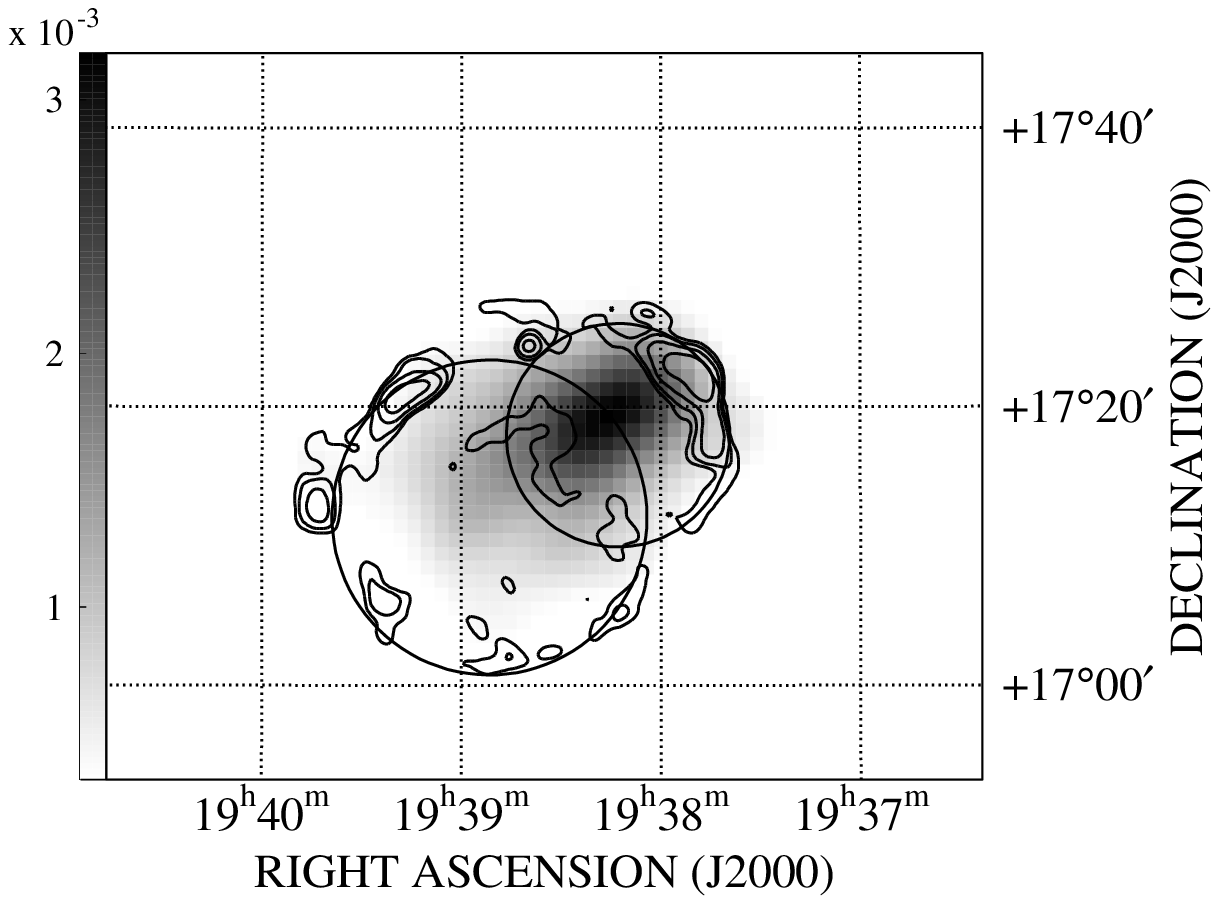}}
   {\footnotesize \setlength{\baselineskip}{9.5pt}%
   Fig.~1.\hspace{4pt}ASCA GIS image of 3C 400.2.
   The contours are the VLA image at 1.4 GHz.
   The contour levels are 0.5, 1, 2, 4, 8 in units of mJy/beam.
   The two circles based on the radio intensity are indicated by solid lines. 
   The circles are $16.\hspace{-2pt}^{\prime}3$ and
   $23.\hspace{-2pt}^{\prime}0$ in diameter.}
 \end{figure*}

We retrieved the ASCA and the ROSAT data of 3C 400.2 from
the HEASARC (High Energy Astrophysics
Science Archive Research Center) public databases.
The ASCA observation was carried out on 1996 April 13 with 
the two Solid-state Imaging Spectrometers (SIS~0 and SIS~1) and
the two Gas Imaging Spectrometers (GIS~2 and GIS~3).
The SIS data were taken in 2-CCD BRIGHT mode whose field of view (FOV) is
$11^{\prime} \times 22^{\prime}$,
while the GISs have a larger FOV of
50$^{\prime}$ in diameter.

Both the GIS data and the SIS data were screened based on
the following criteria.
We rejected events recorded in the South Atlantic Anomaly
(SAA) and the high-background regions with the geomagnetic 
cut-off rigidity of $<$ 6 GV.
The screening criterion for the angle between the rim of the Earth and the
pointing direction was
$< 5^{\circ}$ when the Earth rim is dark, 
and it $< 25^{\circ}$ for the GIS,
$< 40^{\circ}$ for the SIS~0 and
$< 20^{\circ}$ for the SIS~1, when the rim of the Earth is bright.
For the GIS data, we applied a ``flare-cut'' to maximize the 
signal-to-noise ratio, as described in Ishisaki et al.\ (1997).
The SIS data were corrected for the charge-transfer inefficiency
(Dotani et al.\ 1995, 1997), and hot and flickering pixels
were removed using the standard procedures.
The effective exposure times were 19~ks for GIS~2 and GIS~3 and 22~ks for
SIS~0, 24~ks for SIS~1.

We also analyzed the ROSAT PSPC data of 3C 400.2.
We used the basic science data of sequence RP500190N00,
which has already been analyzed by Saken et al.\ (1995).
This observation was performed on 1992 October 4--10 and
the exposure was 3.6~ks.
A summary of the ASCA and ROSAT observations which we analyzed
is given in table~1.

 3C 400.2 is located at $(l, b) =
(53^{\circ}\hspace{-4.5pt}.\hspace{.5pt}6,
-2^{\circ}\hspace{-4.5pt}.\hspace{.5pt}2$), far from 
the galactic center.
In addition, the angular distance from the galactic plane is
larger than the scale height of
the galactic ridge emission of $\sim$ 100~pc
($0^{\circ}\hspace{-4.5pt}.\hspace{.5pt}67$ at 8.5~kpc; Yamauchi et al.\ 1993).
Therefore, we considered the galactic ridge emission to be negligible.
We treated only two components as the GIS background:
the non-X-ray background (NXB)
and the uniform cosmic X-ray background (CXB).
For the GIS,
the NXB data were reproduced using a
method introduced by Ishisaki (1996), and
the CXB were estimated from the Large Sky Survey
(Ueda et al.\ 1999) data.
Total background data consisting of the NXB and CXB were subtracted
from the GIS data.
For the SIS, we considered the blank-sky
(north ecliptic pole and Lynx field regions) data as background.
The background spectrum of the ROSAT PSPC was extracted
from the regions outside the remnant.

We generated exposure-corrected, background-\\
subtracted images of the GIS.
Figure~1 shows a GIS image of 3C 400.2 in gray scale
superposed on the VLA contour map at 1.4~GHz. 
The VLA image was obtained through Skyview
supported by HEASARC/GSFC.
A centrally peaked and elliptical X-ray morphology is clearly seen.

Based on a radio-intensity map,
we delineated the two shells by the regions indicated
by the solid circles in figure~1.
As noted earlier, the shells overlap.
The smaller shell in the NW has a diameter of $16.\hspace{-2pt}^{\prime}3$,
while the larger one in the SE has a diameter of
$23.\hspace{-2pt}^{\prime}0$.
The X-ray surface brightness becomes the maximum inside the NW shell,
which is consistent with previous observations
(Long et al.\ 1991; Saken et al.\ 1995).

\section{Spectral Results}

  \begin{figure}[t]
    \centerline{\psbox[xsize=0.7#1]{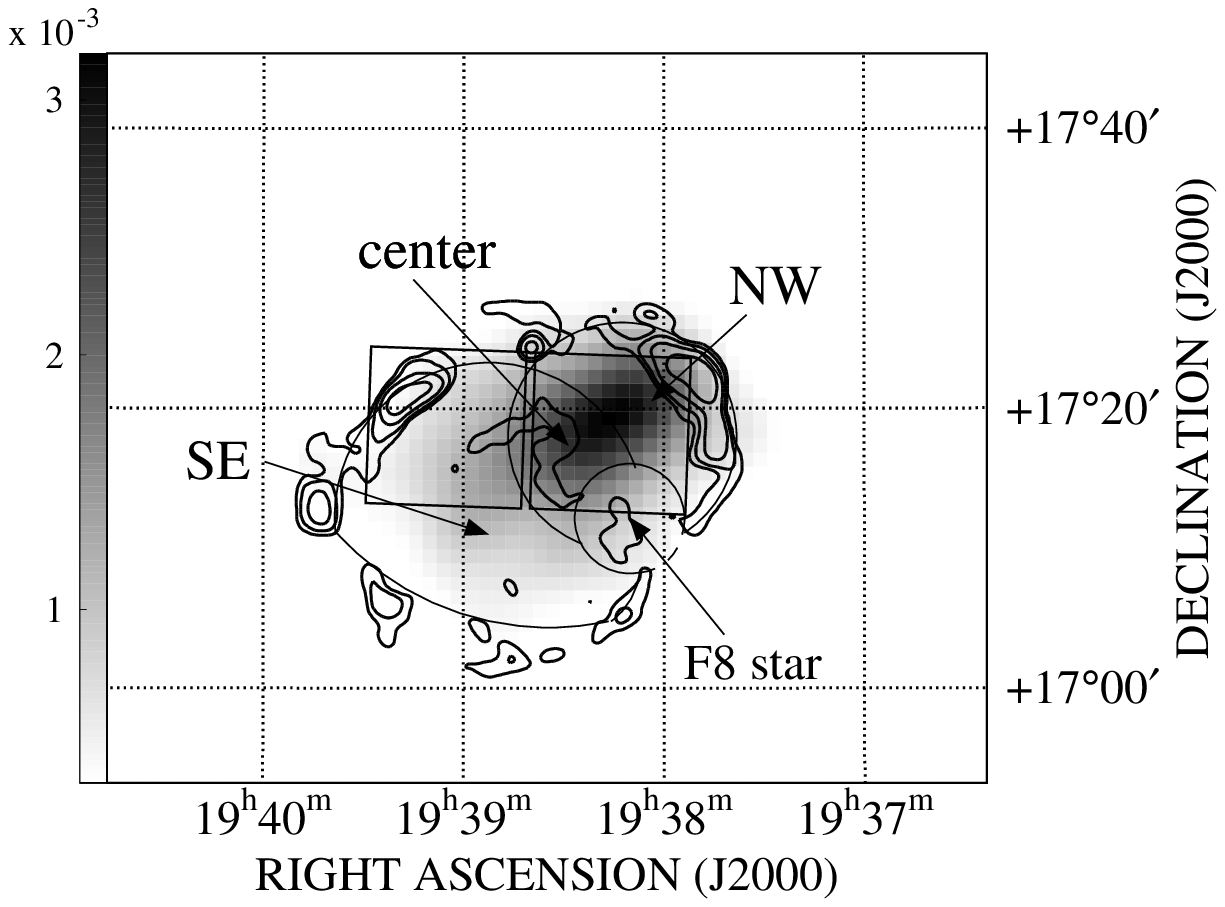}}
    {\footnotesize \setlength{\baselineskip}{9.5pt}%
    Fig.~2.\hspace{4pt}Same as figure~1, but showing various regions
    used in the spectral analysis.
    The thick lines indicate three regions (SE, center, NW) where
    the GIS and PSPC spectra are extracted. The data within the circle in
    the southeast are excluded so as to avoid any contamination by the F8 star
    which was detected with the ROSAT.
    The SIS FOV is indicated by two boxes.}
  \end{figure}
 
We searched for spectral variation over 3C 400.2.
We divided the remnant into three regions:
the overlapping region of two shells (center), and
the remaining parts of the NW shell (NW) and the SE shell (SE).
Figure~2 shows the regions where we extracted the spectra.
The SIS FOV is also shown by the two boxes in figure~2.
They covered only part of the remnant.
Therefore, we cannot use the SIS data for studying the spectral variation.
Since the GIS is sensitive only above 0.7 keV,
we used the ROSAT PSPC data to cover the low-energy band.

There is a point source of an F8 star seen
in the PSPC image of 3C 400.2 (Saken et al.\ 1995)
which is not seen in our GIS data.
We excluded the region of the F8 star
from our spectral analysis in the data screening for the GIS
as well as the PSPC in order to avoid contamination.
In figure~2, the circle, $4^{\prime}$ in diameter, in the southeast
of the remnant indicates the excluded region.

In addition, we excluded the southeastern part of the SE shell.
Because this region corresponds to the edge of the GIS FOV,
the background is high and the gain is uncertain.
We picked up the events within the 21$^{\prime}$ radius from the center of
the GIS FOV.

  \begin{figure}[t]
    \centerline{\psbox[xsize=0.8#1]{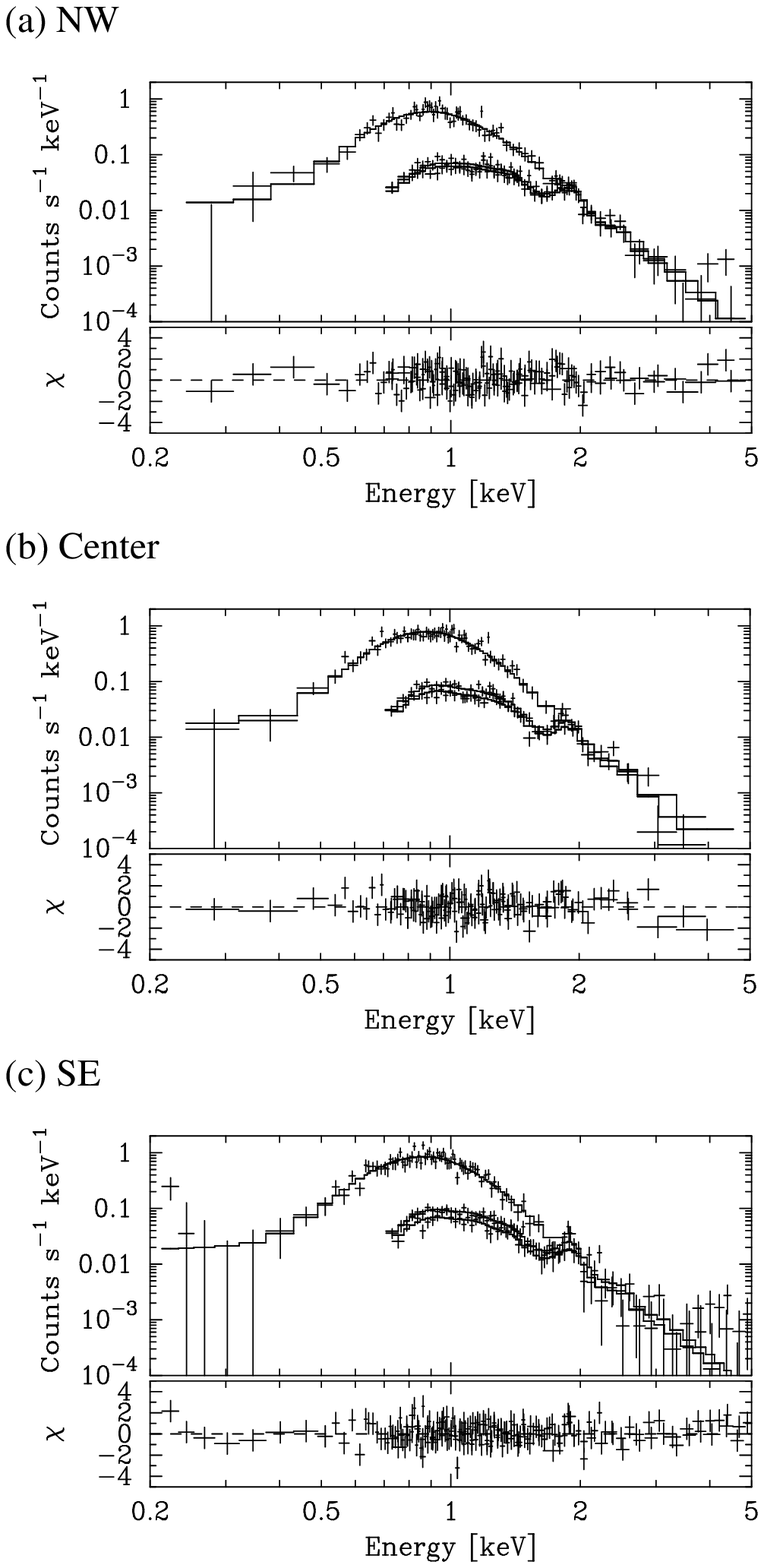}}
    {\footnotesize \setlength{\baselineskip}{9.5pt}%
    Fig.~3.\hspace{4pt}GIS~2, GIS~3, and PSPC spectra
    from three regions of 3C 400.2.
    In each figure, the solid line is the best-fit model of case 3
    (see table 2)
    and the residuals of the fit are shown in the lower panel.}
  \end{figure}%
 
The extracted spectra through this process were binned in such a way
that each channel
contained more than 20 counts to apply the $\chi^{2}$ test.
The background-subtracted spectra are shown in figure~3.
We performed a simultaneous fit
of all three sets of data (GIS~2, GIS~3, and PSPC data)
using XSPEC (V10.0).
The Si lines seen in the spectra indicate
that the X-ray emission surely arises from a hot, thermal plasma,
as had been suggested by previous analyses
(Long et al.\ 1991; Saken et al.\ 1995). 
First, we applied a collisional ionization equilibrium (CIE)
model (MEKAL model, Mewe et al.\ 1985; Liedahl et al.\ 1995)
with cosmic abundances
(Anders, Grevesse et al.\ 1989) to the spectra.
We employed the model of Morrison and McCammon (1983)
as the interstellar absorption feature.
Acceptable fits were obtained in the NW and SE, 
but not for the ``center region'';
$\chi^{2}$/d.o.f = 176.3/138 where d.o.f. stands for the degrees of freedom.
There are large residuals around 1.2 keV which are thought to be due to an
improper fit to the Fe-L line blends.

In order to improve this, we further attempted two cases:
one is to leave the Fe abundance free (Case 1), and the other is to
introduce the non-equilibrium ionization (NEI) model (Case 2).
We employed a VMEKAL model in case 1 and
the NEI model coded by Masai (1984) in case 2, respectively.
The fitting results for the two cases are listed in table~2,
where the errors quoted are at the 90\% confidence level.
Both cases reduced the value of $\chi^{2}$.
In particular, case 2 reduced the $\chi^{2}$ value
for the ``center region''
by $\sim$ 30, which resulted in acceptable fits.
The ionization parameter of log($\tau$) $\sim$ 11 obtained in case 2 
indicates that plasmas have not reached
the ionization equilibrium over the entire remnant.
This is consistent with the low electron density
derived in the following section,
in spite of the old age expected from the optical filaments
and the low electron temperature.

\begin{table*}[t]
  \begin{center}
   Table~2.\hspace{4pt}Best-fit results for each region of 3C400.2. 
The errors are at the 90\% confidence levels.
  \end{center}
  \begin{tabular*}{\textwidth}{@{\hspace{\tabcolsep}
  \extracolsep{\fill}}p{8pc}ccccc}
  \hline\hline \\ [-6pt]
   \hspace*{10pt}Region & $\chi^{2}$/d.o.f &
   $N_{\rm H}$ [$10^{21}{\rm cm}^{-2}$] &
   $kT_{\rm e}$ [keV] & log\,$\tau$ & Fe \\ [4pt]\hline\\[-6pt]
   \multicolumn{6}{l}{Case 1} \\ [2pt]
   \hspace*{10pt}NW \dotfill & 163.4/150 & 4.5$\pm 1.3$ &
   0.56$^{+0.06}_{-0.05}$ & $\cdots$ & 0.8$^{+0.4}_{-0.3}$ \\
   \hspace*{10pt}center \dotfill & 169.1/137 & 4.3$^{+1.1}_{-1.2}$ &
   0.58$^{+0.05}_{-0.06}$ & $\cdots$ & 1.7$^{+0.6}_{-0.4}$ \\
   \hspace*{10pt}SE \dotfill & 169.9/185 & 3.5$\pm 1.2$ &
   0.53$^{+0.06}_{-0.07}$ & $\cdots$ & 1.3$^{+0.5}_{-0.3}$ \\ [6pt]
   \multicolumn{6}{l}{Case 2} \\ [2pt]
   \hspace*{10pt}NW \dotfill & 153.1/150 & 6.2$^{+0.9}_{-1.3}$ &
   0.75$^{+0.16}_{-0.13}$ & 10.7$\pm 0.3$ & $\cdots$ \\
   \hspace*{10pt}center \dotfill & 146.8/137 & 4.7$^{+1.2}_{-2.0}$ &
   0.54$^{+0.12}_{-0.08}$ & 11.2$\pm 0.3$ & $\cdots$ \\
   \hspace*{10pt}SE \dotfill & 158.2/185 & 4.9$^{+0.7}_{-1.4}$ &
   0.62$^{+0.17}_{-0.12}$ & 10.8$^{+0.4}_{-0.3}$ & $\cdots$ \\ [6pt]
   \multicolumn{6}{l}{Case 3} \\ [2pt]
   \hspace*{10pt}NW \dotfill & 153.0/149 & 6.2$^{+0.9}_{-1.3}$ &
   0.76$^{+0.22}_{-0.15}$ & 10.7$\pm 0.3$ & 1.0$^{+0.4}_{-0.3}$ \\
   \hspace*{10pt}center \dotfill & 134.9/136 & 4.4$^{+1.5}_{-1.3}$ &
   0.80$^{+0.23}_{-0.20}$ & 10.9$^{+0.3}_{-0.2}$ & 2.0$^{+0.6}_{-0.5}$ \\
   \hspace*{10pt}SE \dotfill & 154.8/184 & 4.8$^{+0.8}_{-1.4}$ &
   0.83$^{+0.44}_{-0.24}$ & 10.6$\pm 0.3$ & 1.4$^{+0.6}_{-0.4}$ \\ [4pt]
   \hspace*{10pt}SIS FOV \dotfill & 493.8/465
   & 3.2$^{+0.4}_{-0.5}$, 2.1$\pm 0.3^\ast$ &
   0.76$^{+0.05}_{-0.04}$ & 11.2$\pm 0.1$ & 1.5$\pm 0.2$ \\ [4pt]
   \hline
  \end{tabular*}
 \vspace{6pt}\par\noindent
 $\ast$ Extra $N_{\rm H}$ for the SIS.
\end{table*}

Finally, we applied the NEI model, allowing the abundance of
Fe to be free parameters (Case 3).
The best-fit results summarized in table~2 are shown
by the solid lines in figure~3.
We found no significant variation of any
parameters, column density ($N_{\rm H}$),
electron temperature ($kT_{\rm e}$), ionization parameter [log($\tau$)],
within the statistical uncertainties over the entire remnant.
The abundance of Fe shows a slight enhancement in the ``center region''.

\begin{table*}[t]
  \begin{center}
   Table~3.\hspace{4pt}
   Emission measure ($n_{\rm e}^2L$) for each region of 3C400.2.
  \end{center}
  \begin{tabular*}{\textwidth}{@{\hspace{\tabcolsep}
  \extracolsep{\fill}}p{10pc}ccc}
  \hline\hline \\ [-6pt]
   & NW & Center & SE \\ [4pt]\hline\\[-6pt]
   $n_{\rm e}^2L$ [${\rm cm}^{-6}$ pc] \dotfill
   & 0.21$\pm 0.03$ & 0.14$\pm 0.02$ & 0.058$^{+0.006}_{-0.008}$ \\
   $n_{\rm e}^2L$/$(n_{\rm e}^2L)_{\rm SE}$ \dotfill
   & 3.6$\pm 0.6$ & 2.5$\pm 0.5$ & (1.0) \\ [4pt]
   \hline
  \end{tabular*}
\end{table*}

Table~3 summarizes the emission measures ($n_{\rm e}^2L$ where $L$ is
the depth of the plasma)
for three regions which are calculated 
using the results for GIS in case 3.
We fixed the $kT_{\rm e}$ value when we derived
the error range of the $n_{\rm e}^2L$.
The ratio of $n_{\rm e}^2L$ between SE and other regions is also
given in table~3.
The $n_{\rm e}^2L$ value tends to be larger from the southeast
to the northwest.

  \begin{figure}[t]
    \centerline{\psbox[xsize=0.72#1]{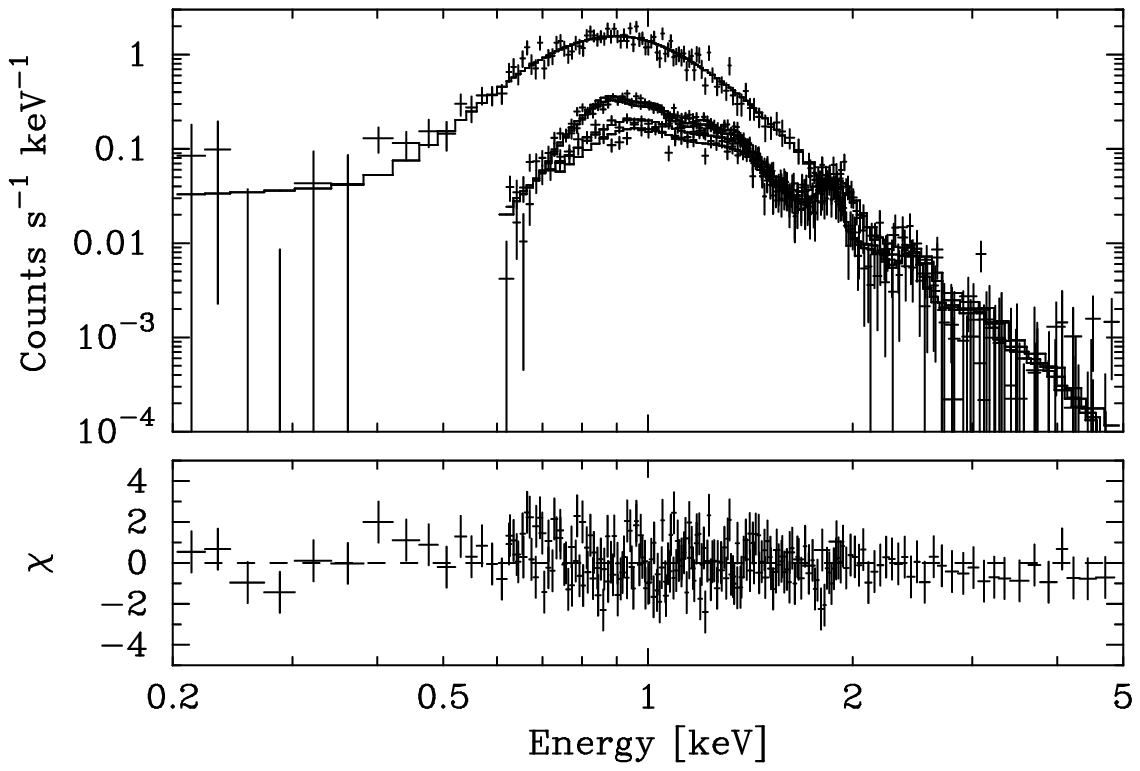}}
    {\footnotesize \setlength{\baselineskip}{9.5pt}%
    Fig.~4.\hspace{4pt}Top panel showing the SIS~0, SIS~1, GIS~2, GIS~3, and
    PSPC spectra extracted
    from the SIS FOV. The lower panel shows the residuals of the fit for only
    SIS~0 and PSPC.
    The solid line is the best-fit model of case 3 (see table 2).}
  \end{figure}
 
We also performed the spectral fitting with the model of case 3
including the SIS data. 
We have extracted the spectra for three instruments
(SIS, GIS, and PSPC) from the region corresponding to the SIS FOV.
The obtained spectra are shown in figure~4.
Since it has been reported that SIS spectra 
tend to show a higher value of the column density 
(ASCA Guest Observer Facility's home page at 
{\it http://heasarc.gsfc.nasa.gov/docs/asca/cal\_probs.html}; Orr et
al.\ 1998), we added an extra absorption feature for the SIS to
the model of case 3.
The results are listed in the bottom of table~2, and
are consistent with those of the GIS/PSPC spectral fitting.

\section{Discussion}

As mentioned above,
3C 400.2 shows a unique morphology like two overlapping shells 
in radio.
The mechanism which produces this morphology has been
investigated in previous papers
(Dubner et al.\ 1994; Saken et al.\ 1995; Giacani et al.\ 1998).
We discuss it based on our results of the spectral fitting.
The most interesting hypothesis is that 3C 400.2 consists of two
interacting SNRs.

\begin{table*}[t]
  \begin{center}
   Table~4.\hspace{4pt}
   Parameters of 3C 400.2 and other mixed-morphology SNRs.   
  \end{center}
  \begin{tabular*}{\textwidth}{@{\hspace{\tabcolsep}
  \extracolsep{\fill}}p{8pc}cccc}
  \hline\hline \\ [-6pt]
   & 3C~400.2 & 3C~391 & W~44 & W~28 \\  [4pt]\hline\\[-6pt]
   Size [pc$^{2}$] \dotfill & ($44\times 35$) $D^2_5$\hspace{2pt}$^{\ast, \dagger}$
   & $21\times 18^\ddagger$ & $29\times 17${\scriptsize $^{^{\S}}$}
   & $24\times 24${\scriptsize $^{^{\P}}$}\\
   $kT_{\rm e}$ [keV] \dotfill & 0.8$^{+0.2}_{-0.1}$
   & 0.5$^\ddagger$ & 0.88{\scriptsize $^{^{\S}}$}
   & 2.0{\scriptsize $^{^{\P}}$} \\
   $n_{\rm e}$ [cm$^{-3}$] \dotfill & (0.05$\pm 0.01$) $D^{-1/2}_5$
   & 0.63$^\ddagger$ & 0.42{\scriptsize $^{^{\S}}$}
   & 0.13{\scriptsize $^{^{\P}}$}\\
   $E_{\rm t}$ [erg] \dotfill & (5.6$\pm 1.4$) $D^{5/2}_5$ $\times 10^{49}$
   & 6.1$\times 10^{49}$ & 7.7$\times 10^{49}$
   & 2.6$\times 10^{50}${\scriptsize $^{^{\P}}$}\\
   $M_{\rm X}$ [$M_{\odot}$] \dotfill & (38$\pm 7$) $D^{5/2}_5$
   & 74$^\ddagger$ & 56{\scriptsize $^{^{\S}}$}
   & 27{\scriptsize$^{^{\P}}$} \\ [4pt]
   \hline
  \end{tabular*}
 \vspace{6pt}\par\noindent
 $\ast$ $D_5$ is the distance in units of 5~kpc. \\
 $\dagger$ $30^{\prime}\times 24^{\prime}$ (Saken et al.\ 1995) \\
 $\ddagger$ Rho and Petre (1996) \\
 {\small {\S}} Harrus et al.\ (1997) \\
 {\small {\P}} Long et al.\ (1991)
\end{table*}

Bodenheimer et al.\ (1984) has carried out two-dimensional numerical
calculations of the interacting of two SNRs in a uniform-density
interstellar medium (ISM) with 1 cm$^{-3}$.
They performed calculations while varying the time interval and
distance between the two supernova (SN) explosions.
The morphology, which resembles that of 3C 400.2, was obtained 
in the case that the second SN explodes in the expanding shell
of the first SN.
The second SN explosion occurs 1.37$\times$10$^4$ yr after the first SN.
They calculated the X-ray intensity distribution at 1~keV, which
shows a centrally peaked and non-spherical morphology
at 8.7$\times$10$^4$ yr after the first SN.
Furthermore, they presented the evolution of the density and the
temperature.
Their simulation indicates a large temperature variation of $\sim10^4$ 
from the center toward the shell.
However, our analysis of 3C 400.2 indicates no variation of the temperature
within the X-ray emitting region.
Therefore, we conclude that 
the interaction of two SNRs does not reproduce the case of 3C 400.2. 
Saken et al.\ (1995) drew the same conclusion
from the lack of a large temperature variation across the remnant
with only the ROSAT PSPC data.

In addition, we investigated the possibility that the morphology of
3C 400.2 arises from two SNRs which are not interacting, but which are
projected along the same line of sight.
In this case, $n_{\rm e}^2L$ in the region where two shells are
overlapping would become the sum of
$(n_{\rm e}^2L)_{\rm NW}$ and $(n_{\rm e}^2L)_{\rm SE}$.
At least, we would expect that the ``center region'' has a larger
value of $n_{\rm e}^2L$ than other regions.
Nevertheless, we could find no evidence of an excess value of
$n_{\rm e}^2L$ in the ``center region'', as shown in table~3. 
The hypothesis of the projection effect cannot apply to 3C 400.2.

Since it is unlikely that 3C 400.2 consists of two SNRs,
whether they are interacting or not,
we consider the case that it is
a single SNR.\ \ Rho and Petre (1998) proposed that
3C 400.2 belongs to a class of ``mixed-morphology SNRs''.
They distinguish ``mixed-morphology SNRs'' from others with the
following criteria.
They are classified as a shell-type SNR at radio wavelength,
while the X-ray morphology is
centrally peaked or amorphous.
The X-ray emission is thermal emission from the ISM, not 
from the ejecta.
In addition, there is no prominent,
central, compact source in radio and X-ray bands.
Based on these criteria,
they give 7 SNRs as prototypical mixed-morphology SNRs.
Besides 3C 400.2,
it includes W~28, W~44, Kes~27, MSH~11$-$61A, 3C~391, and CTB~1.

In order to obtain the mean value of parameters over 3C 400.2,
we performed a spectral fitting of the summed data
of three regions for the GIS and the PSPC.
The condition of the spectral fitting is the same as in case~3 for each region.
The obtained best-fit values are consistent with those given in table~2.
Using these values,
we calculated the electron density ($n_{\rm e}$), 
the thermal energy ($E_{\rm t}$), and
the mass of X-ray emitting material ($M_{\rm X}$).
The remnant is assumed to be an ellipsoid with two axes of 
30$^{\prime}$ and 24$^{\prime}$ in the plane of
the sky (Saken et al.\ 1995), and
the depth which is equal to
the length of a minor axis, 24$^{\prime}$.
We also assumed the volume-filling factor to be unity and 
that the mean number density of hydrogen is equal to that of electrons,
for simplicity.
The results are given in table~4,
where $D_5$ is the distance in units of 5~kpc.
In addition, we have also listed the parameters
for three other remnants among prototypical mixed-morphology SNRs,
3C~391, W~44 and W~28.
At the writing phase of this paper,
we could pick up the parameters of these three SNRs from the literature
in order to compare 3C 400.2.
We excluded CTB 1 from our list, since it has a complicate structure,
as mentioned in section~1.
The properties of 
3C 400.2 are similar to those of 3C~391 and W~44,
except for the electron density.
The higher $kT_{\rm e}$ and the larger $E_{\rm t}$ of W~28
may be due to the contribution of non-thermal emission detected with
the ASCA observation (K. Torii 1995, private communication).

A centrally peaked X-ray morphology observed in
mixed-morphology SNRs cannot be derived
from the Sedov (1959) similarity solution
for a point explosion in a uniform-density ISM.\ \
Rho and Petre (1998) gave two most likely models which can explain 
the properties of mixed-morphology SNRs.
One is a model in which the plasma at a shell has cooled to below
$\sim 10^6$ K and its emission cannot be detectable in X-rays.
The interior of the remnant is still hot due to a lower density
than that at the shell.
In this case, the plasma inside the shell must come from
the ejecta, just as observed in the Cygnus Loop (Miyata et al.\ 1998).
However, it is inconsistent with cosmic abundances
observed across 3C 400.2.

An alternative explanation is the cloud evaporation model
(White, Long 1991 and references therein).
In this model, it is assumed that
there are small and numerous cloudlets embedded in the ISM.
These clouds have a much higher density
than that of the tenuous and hot ISM.
Therefore, they would not be accelerated
by the passage of a shock wave.
The clouds remaining in the interior of the shell gradually
evaporate and emit X-rays.
The thermal energy listed in table~4 represents the energy 
given to clouds by a shock wave.
Long et al.\ (1991) found that
the Einstein observation of 3C 400.2 can be
explained by this model.
We note that the important point in the cloud evaporation model 
is the presence of numerous, small and
dense cloudlets in the environment of a SNR.

Among mixed-morphology SNRs,
the most interesting object is 3C 391, with respect to
the similarity to 3C 400.2.
3C 391 shows an elliptical shape with elongation from the northwest
to the southeast, which resembles the morphology of 3C 400.2.
In the radio band, a ``breakout'' structure is also seen in the
southeast (Reynolds, Moffett 1993).
Based on the radio and X-ray observations,
Rho and Petre (1996) suggest that 3C 391 is the result of an explosion
near to the edge of the molecular cloud.
When the shock wave reaches the cloud boundary,
a breakout would occur.
Moreover, it is likely that there are numerous, small and
dense cloudlets, which are required in the cloud evaporation model,
at the edge of the molecular cloud.
The CO observations confirmed the association between 3C 391 and
a molecular cloud (Wilner et al.\ 1998).

In the case of 3C 400.2,
Giacani et al.\ (1998) identified a dense H\,{\small I} cloud,
which they argued is associated with 3C 400.2.
In particular, the H\,{\small I} cloud surrounds the north and western side of
the radio shell in the velocity range +16.5 to +28 km$^{-1}$.
As Giacani (1998) proposed,  
a shock front propagating toward the southeast would
meet a lower density medium and form the large shell by a breakout.
Therefore, the morphology of 3C 400.2 can also result from
an explosion occurring near to the edge of the H\,{\small I} cloud,
as well as the case of 3C 391.
This scenario can explain both the elongated radio morphology and
the centrally peaked X-ray emission.
As shown in table~2,
the larger value of $N_{\rm H}$ in the NW region may indicate
the density gradient, though it is not clear due to large errors.
Rho and Petre (1998) reported that
most of mixed-morphology SNRs are interacting with
molecular or H\,{\small I} clouds.
Follow-up observations, such as a detailed CO mapping in the vicinity of
3C 400.2 or the detection of shock-excited OH maser emission, would be
useful to confirm the interaction between 3C 400.2 and the cloud.

\section{Conclusion}

We performed a spectral analysis of 3C 400.2 with the ASCA and
the ROSAT PSPC data.
For all regions of 3C 400.2,
we obtained a good fit by employing the NEI plasma model with cosmic
abundances, except for a slight enhancement of Fe.
No significant variation in the electron
temperature and the ionization parameter
was found across the entire remnant with $kT_{\rm e} \sim 0.8$ keV and
log\,$\tau$ $\sim$ 11.
On the other hand, the emission measure ($n_{\rm e}^2L$) tends to be
larger from the southeast to the northwest.
We therefore conclude that 3C 400.2 does not consist of two SNRs;
nevertheless, the radio morphology of 3C 400.2 seems to be
an overlapping of two shells. 

We compared the values of $kT_{\rm e}$,
$n_{\rm e}$, $E_{\rm t}$, and $M_{\rm X}$ between 3C 400.2 and other 
prototypical mixed-morphology SNRs.
Although $n_{\rm e}$ for 3C 400.2 is smaller than those of 3C~391 and W~44,
other parameters show similar values.
Based on these similarities and the existence of a
dense H\,{\small I} cloud, we conclude that the morphology of 3C 400.2 is most
easily explained in terms of a supernova explosion near to the edge of the
cloud and a cloudlet evaporation model for the SNR.
\par
\vspace{1pc}\par
We are grateful to all the members of the ASCA team. 
K.Y.\ is supported by JSPS Research Fellowship for Young Scientists.

\clearpage
\section*{References}

\re
Agrawal, P.C., Riegler, G.R., \& Singh, K.P.\ 1983, Ap\&SS, 89, 279

\re
Allakhverdiev, A.O., Guseinov, O.Kh., Kasumov, F.K., \& Iusifov, I.M.\ 1983,
Ap\&SS, 97, 287

\re
Anders, E., \& Grevesse, N.\ 1989, Geochim.\ Cosmochim.\ Acta, 53, 197

\re
Bodenheimer, P., Yorke, H.W., \& Tenorio-Tagle, G.\ 1984, A\&A, 138, 215

\re
Case, G.L., \& Bhattacharya, D.\ 1998, ApJ, 504, 761 

\re
Caswell, J.L., \& Lerche, I.\ 1979, MNRAS, 187, 201 

\re
Clark, D.H., \& Caswell, J.L.\ 1976, MNRAS, 174, 267

\re
Craig W.W., Hailey C.J., \& Pisarski R.L.\ 1997, ApJ 488, 307

\re
Dotani, T., Yamashita, A., Ezuka, H., Takahashi, K., Crew, G., Mukai, K., \& 
the SIS team 1997, ASCA News, 5, 14 

\re
Dotani, T., Yamashita, A., Rasmussen, A., \& the SIS team 1995,
ASCA News, 3, 25 

\re
Dubner, G.M., Giacani, E.B., Goss, W.M., \& Winkler, P.F.\ 1994, ApJ, 108, 207

\re
Giacani, E.B., Dubner, G., Cappa, C., \& Testori, J.\ 1998, A\&AS, 133, 61

\re
Goss, W.M., Siddesh, S.G., \& Schwartz, U.J.\ 1975, A\&A, 43, 459

\re
Harrus, I.M., Hughes, J.P., Singh, K.P., Koyama, K., \& Asaoka, I.\ 1997, ApJ, 488, 781

\re
Ishisaki, Y.\ 1996, Ph.D.\ thesis, The University of Tokyo

\re
Ishisaki, Y., Ueda, Y., Kubo, H., Ikebe, Y.,
Makishima, K., \& the GIS team 1997, ASCA News, 5, 26 

\re
Liedahl, D.A., Osterheld, A.L., \& Goldstein, W.H.\ 1995, ApJ,
438, L115

\re
Long, K.S., Blair, W.P., White, R.L., \& Matsui, Y.\ 1991, ApJ, 373, 567

\re
Masai, K.\ 1984, Ap\&SS, 98, 367

\re
Mewe, R., Gronenschild, E.H.B.M., \& 
van den Oord, G.H.J.\ 1985, A\&AS, 62, 197 

\re
Milne, D.K.\ 1979, Austral.\ J.\ Phys.\, 32, 83

\re
Miyata, E., Tsunemi, H., Kohmura, T., Suzuki, S., \& Kumagai, S.\
1998, PASJ, 50, 257

\re
Morrison, R., \& McCammon, D.\ 1983, ApJ, 270, 119

\re
Nishiuchi, M., Yokogawa, J., Koyama, K., \& Hughes, J. P.\ 2001, PASJ, 53 in press

\re
Orr, A., Yaqoob, T., Parmar, A.N., Piro, L., White, N.E., \& Grandi, P.\ 1998, A\&A, 337, 685 

\re
Reynolds, S.P., \& Moffett, D.A.\ 1993, AJ, 105, 2226

\re
Rosado, M.\ 1983, Rev.\ Mexicana Astron.\ Astrofis., 8, 59

\re
Rosado, M., Laval, A., Boulesteix, J., Georgelin, Y.P., Greve, A.,
Marcelin, M., Coarer, E.Le, \& Viale, A.\ 1990, A\&A, 238, 315

\re
Rho, J.-H., \& Petre, R.\ 1996, ApJ, 467, 698

\re
Rho, J.-H., \& Petre, R.\ 1998, ApJ, 503, L167

\re
Saken, J.M., Long, K.S., Blair, W.P., \& Winkler, P.F.\ 1995, ApJ, 443, 231

\re
Sedov, L.I.\ 1959, Similarity and Dimensional Methods in Mechanics
(New York: Academic Press)

\re
Ueda, Y., Takahashi, T., Inoue, H., Tsuru, T., Sakano, M., Ishisaki, Y.,
Ogasaka, Y., Makishima, K.\ et al.\ 1999, ApJ, 518, 656 

\re
White, R.L., \& Long, K.S.\ 1991, ApJ, 373, 543

\re
Williams, R.M., Chu, Y-H., Dickel, J.R., Beyer, R., Petre, R.,
Smith, R.C., \& Milne, D.K.\ 1997, ApJ, 480, 618

\re
Williams, R.M., Chu, Y-H., Dickel, J.R., Petre, R.,
Smith, R.C., \& Tavarez, M.\ 1999, ApJS, 123, 467

\re
Wilner, D.J., Reynolds, S.P., \& Moffett, D.A.\ 1998, AJ, 115, 247

\re
Winkler, P.F., Olinger, T.M., \& Westerbeke, S.A.\ 1993, ApJ, 405, 608

\re
Yamauchi, S., \& Koyama, K.\ 1993, ApJ, 404, 620 

\re
Yoshita, K., Miyata, E., \& Tsunemi, H.\ 2000, PASJ, 52, 867

\end{document}